\documentclass[twocolumn,showpacs,preprintnumbers,amsmath,amssymb,a4paper]{revtex4}

\usepackage{amsmath,amssymb,bbm,epsfig}

\newcommand{\be}{\begin{equation}}
\newcommand{\ee}{\end{equation}}
\newcommand{\ba}{\begin{array}}
\newcommand{\ea}{\end{array}}
\newcommand{\bqa}{\begin{eqnarray}}
\newcommand{\eqa}{\end{eqnarray}}

\newcommand{\tr}{\mbox{Tr}}
\newcommand{\bra}[1]{\ensuremath{\langle #1 |}}
\newcommand{\ket}[1]{\ensuremath{| #1 \rangle}}

\begin{document}

\title{Robust entangled states}

\author{Florian Mintert}

\affiliation{
Physikalisches Institut, Albert-Ludwigs Universit\"at Freiburg,
Hermann-Herder-Str. 3, Freiburg, Germany}

\date{\today}

\begin{abstract}
We establish a technique to find the states with most robust entanglement in dissipative quantum systems
and explicitly construct those state for various environments.
\end{abstract}


\maketitle

The potential to form entangled states is one of the central distinctions between quantum objects and their classical counterparts.
Therefore, the stability properties of entanglement teaches us a lot about the emergence of classicality in quantum systems
of growing size.
Besides that -- or also, even more importantly -- entangled states are the central building block for many promising applications \cite{raussendorf:5188}
in quantum information theory,
so that there is a considerable interest not only in the preparation of highly entangled states,
but also in the preservation of entanglement over sufficiently long times that allow to execute a quantum algorithm or different tasks. 

Generally, entanglement decays due to environment coupling \cite{dur:180403,carvalho:230501,aolita:080501}.
However, the decay of entanglement does not necessarily follow that of the density matrices coherences.
In particular, since entanglement is {\em not} a linear function of the underlying quantum state,
there can be states whose entanglement is significantly more robust then that of other states.

Here, we seek those states whose entanglement is most robust in a given situation of environment coupling.
That is, given an entanglement measure $E$, we look for those states, for which the temporal increment $\dot E$ is maximal.

Such tasks require an entanglement measures that allows a simple evaluation or estimation.
Entanglement measures, however, are rather intricate to evaluate since most of them rely on a mathematical optimization procedure
that can be solved without extensive numerical optimizations only in exceptional cases.
What we consider here, is not a measure $E(\varrho)$ itself, but rather its time derivative $\dot E(\varrho,\dot\varrho)$,
and the technical difficulties to evaluate the letter quantity are certainly not smaller than for the former.

\begin{figure*}
\parbox{\textwidth}{
\includegraphics[width=0.3\textwidth,angle=0]{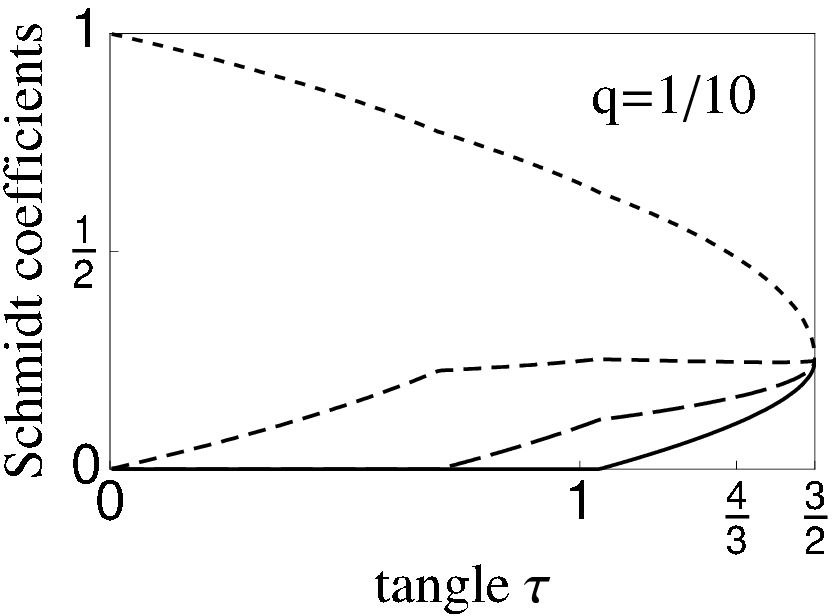}\hspace{0.04\textwidth}
\includegraphics[width=0.3\textwidth,angle=0]{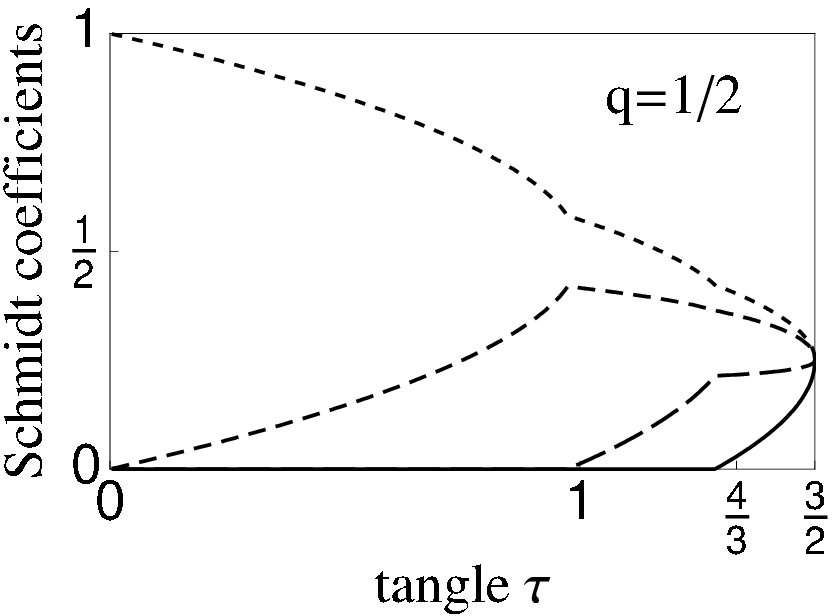}\hspace{0.04\textwidth}
\includegraphics[width=0.3\textwidth,angle=0]{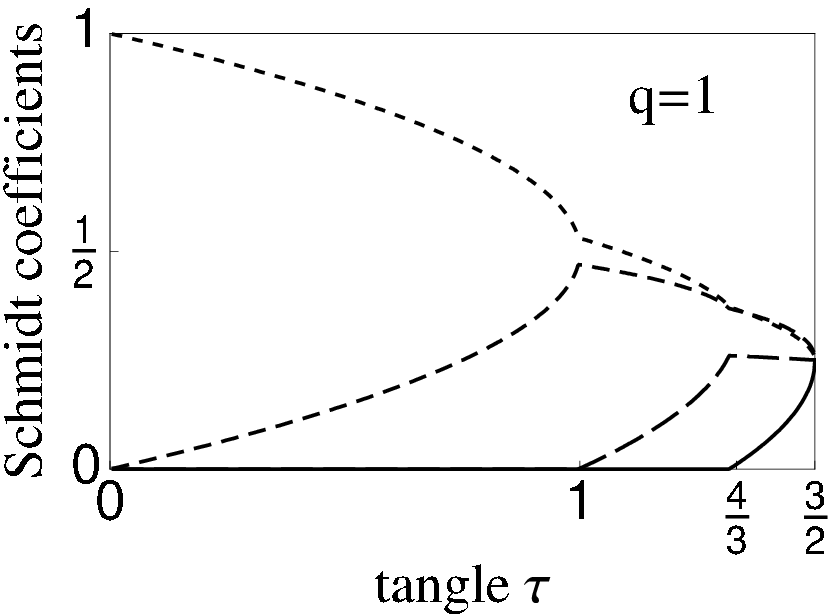}}
\caption{Schmidt coefficients of the states with the most robust entanglement
against spontaneous decay as function of tangle $\tau$.
The different insets correspond to different values of the parameter $q$ that characterizes the decay mechanism in Eq.~\eqref{decayop}.
The Schmidt coefficients are distinguished by their dashing with increasing length of dashing from $\lambda_0$ (shortest dashing) to $\lambda_3$ (solid line).}
\label{lambda_decay}
\end{figure*}

Therefore, we make use of a purely algebraic estimate of concurrence \cite{PhysRevA.54.3824} in terms of a bilinear functional of the density matrix $\varrho$.
The concurrence of a pure state $\ket{\Psi}$ can be defined via a spin-flip operation for two-level systems \cite{PhysRevA.54.3824},
or in terms of a linearized entropy of the reduced density matrix of one of the subsystems \cite{PhysRevA.64.042315},
or, also, via the expectation value of a suitably chosen operator with respect to the duplicate quantum state $\ket{\Psi}\otimes\ket{\Psi}$ \cite{mintert:260502}
for systems of arbitrary finite dimension.
The extension to mixed states can be performed with the help of a convex roof \cite{uhl97}
$c(\varrho)=\inf\sum_ip_i\ket{\Psi_i}\bra{\Psi_i}$, where the infimum is to be found among all pure state decompositions
$\varrho=\sum_ip_i\ket{\Psi_i}\bra{\Psi_i}$ of the state $\varrho$.
Whereas this infimum can be found algebraically only for small systems \cite{PhysRevLett.80.2245} or some special states \cite{ter00}, 
it can be bounded from below by
\be
c^2(\varrho)=\tau(\varrho)\ge\tr \varrho\otimes\varrho V\ ,
\label{tau}
\ee 
for general states \cite{mintert:140505},
where $V=P_-\otimes P_- -1/2(P_-\otimes P_+ + P_+\otimes P_-)$
is defined in terms of the projectors $P_\mp$ on the antisymmetric and symmetric components of the duplicated Hilbert spaces of either subsystem,
and $\tau=c^2$ is called tangle.
This is exact for pure states, and provides a very good approximation for weakly mixed states,
what makes it an ideal tool for the following investigations in which we will consider the decay of initially pure states into mixed ones.
With this bound, we can approximate the temporal increment of tangle as
\be
\dot \tau(\varrho,\dot\varrho)\simeq 2\tr(\dot\varrho\otimes\varrho V)\ ,
\label{taudot}
\ee
what allows to maximize $\dot\tau$ over pure initial states.
Typically, weakly entangled states show more robust entanglement than highly entangled ones,
and initially separable states have a vanishing temporal increment under coupling to local environments.
In order to ensure that our optimizations will not simply yield separable states as those with most robust entanglement,
we will fix the initial tangle $\tau(\Psi)=\tau_0$,
and seek those states with given $\tau_0$ that maximize the temporal increment of tangle
\be
\dot\tau_{\mbox{opt}}=\max\left(\dot\tau(\Psi) | \tau(\Psi)=\tau_0\right)\ .
\ee

Environment coupling will be described in the following in terms of
a Master equation \cite{Lindblad:1976fj,klh}
$\dot\varrho=\sum_i{\cal L}_{\sigma_i}\varrho$,
with a Lindbladian ${\cal L}_{\sigma}$ that acts like
\be
{\cal L}_{\sigma}\varrho=\left(2\sigma\varrho\sigma^\dagger-\sigma^\dagger\sigma\varrho-\varrho\sigma^\dagger\sigma\right)\ .
\label{lindblad}
\ee
For coupling operators $\sigma$ we consider the case of {\em local} environments,
where any such quantities acts nontrivially only on a single subsystem,
since this corresponds to the typical situation in which the subsystems are macroscopically separated.
The two iconic models for evolutions of quantum states in quantum information processing are the the `amplitude damping' channel
that corresponds to $\sigma=\sigma_-=(\sigma_x-i\sigma_y)/2$,
and the `phase damping' channel that corresponds to $\sigma=\sigma_z$ \cite{chuang00} in terms of the Pauli spin operators.
Here, we consider generalizations of these situations to quantum systems with more than just two levels.
Typical generalizations to higher dimensional systems are often derived for the harmonic oscillator,
where the generalization of $\sigma_-$ is the annihilation operator $a$,
and the generalization of $\sigma_z$ is the number operator $n=a^\dagger a$.
This, however, is a rather specific situation that can not describe general experimental situations sufficiently well.
For example, there could be several excited states that have comparable life time, or rather different ones that are incompatible with those
resulting from the harmonic oscillator operators. 
We therefore take the generalization
\be
\sigma_{\mbox{dc}}(q)=\sum_i (i+1)^q\ket{i}\bra{i}\ ,
\label{decoherenceop}
\ee
of the harmonic oscillator case to describe decoherence.
Here, there is the variable parameter `$q$' that allows to change the properties of the dephasing mechanism.
The harmonic oscillator case is recovered for $q=1$,
but changing the value of $q$ allows to adjust the decay rates for the coherences between different levels.
As generalization of spontaneous decay, we take
\be
\sigma_{\mbox{ad}}(q)=\sum_i (i+1)^q\ket{0}\bra{i+1}\ ,
\label{decayop}
\ee
that is the case where all excited levels decay to the ground state with adjustable decay rates.
This is different to the harmonic oscillator case, where excited states decay to the ground state via all lower lying states,
but rather resembles the situation of an atom or ion with several excited states the decay directly to the ground state.
 
Obviously, also these models can not provide an exhaustive description of very general dissipative dynamics.
However, these two situations will provide a good insight in the interplay of stability of entanglement and environment coupling,
and situations with specific differences as compared to Eqs.~\eqref{decoherenceop} and \eqref{decayop} can easily be investigated along the lines
that we present in the following. 

Like any entanglement measure, the tangle is invariant under local unitaries,
and given this invariance, any state can be represented in its Schmidt decomposition \cite{schmidt_deco}
\be
\sum_i\sqrt{\lambda_i}\ket{\phi_i}\otimes\ket{\bar\phi_i}\ ,
\label{Schmidt}
\ee
where all entanglement properties are described in terms of the Schmidt coefficients $\lambda_i$,
and $\{\ket{\phi_i}\}$ and $\{\ket{\bar\phi_i}\}$ are local bases for the individual subsystems. 
However, the environment coupling typically breaks this invariance for $\dot\tau$,
so that the present maximization can not be restricted to states that are of Schmidt form (Eq.~\eqref{Schmidt}).
Nevertheless, most situations of environment coupling define a special basis,
the so-called pointer basis that consists of states that are most robust under environment coupling.
Performing numerical optimizations over general states we found that typically states with such pointer bases as Schmidt bases
have the most robust entanglement.
We will therefore take exactly those states, optimize over their Schmidt coefficients,
and convince ourselves at the end with comparison to numerical optimizations over general states that this restriction still
allows to find the states with the most robust entanglement.

Given an initial state in Schmidt decomposition, the most robust entanglement is determined by the solution of polynomial equations
\be
\sum_i\alpha_i\frac{\partial \dot\tau(\Psi)}{\partial \lambda_i}=0\ , \ \mbox{and}\
\tau(\Psi)=\tau_0
\ee
where the prefactors $\alpha_i$ need to satisfy $\sum_i\alpha_i\frac{\partial \tau(\Psi)}{\partial \lambda_i}=0$
to ensure that the variation over the $\lambda_i$ is performed over states with a constant tangle.
Such sets of polynomial equations can very reliably and efficiently be solved with the help of Groebner bases \cite{groebner}.
The risk of finding local instead of global minima that always affects numerical optimization is basically dispelled.

\begin{figure*}
\parbox{\textwidth}{
\includegraphics[width=0.24\textwidth,angle=0]{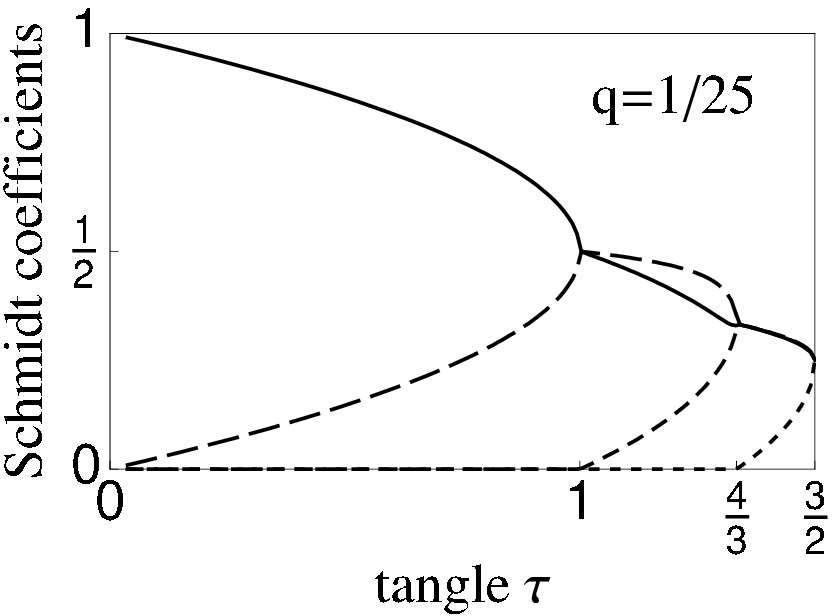}
\includegraphics[width=0.24\textwidth,angle=0]{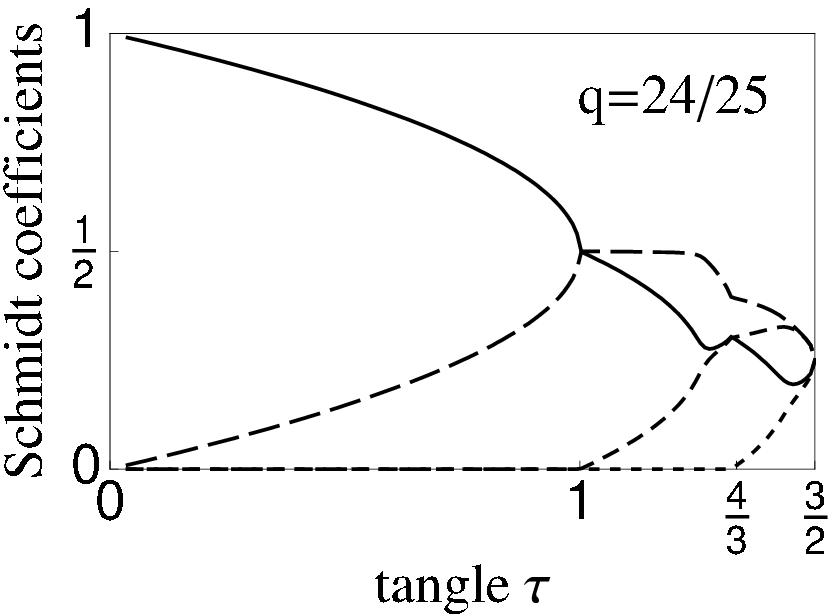}
\includegraphics[width=0.24\textwidth,angle=0]{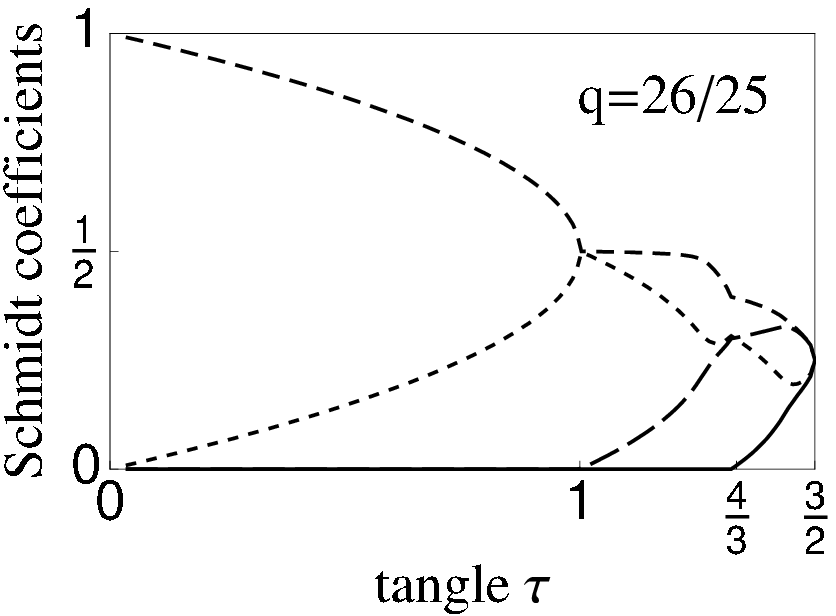}
\includegraphics[width=0.24\textwidth,angle=0]{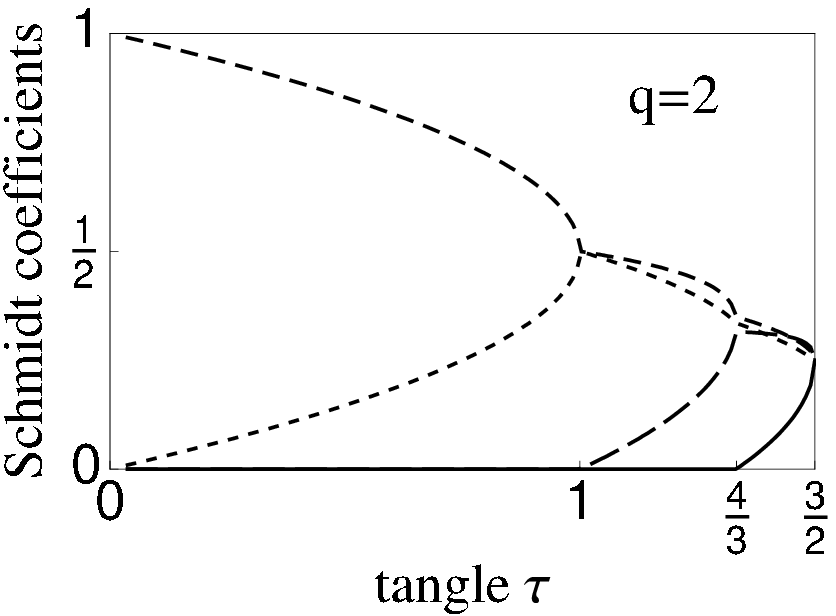}}
\caption{Schmidt coefficients of the states with most robust entanglement against decoherence analogous to Fig.~\ref{lambda_decay}.
The different insets correspond to different values of $q$ defined in Eq.~\eqref{decoherenceop}.}
\label{lambda_decoherence}
\end{figure*}

Fig.~\ref{lambda_decay} shows the Schmidt coefficients of the states with the most robust tangle against environment coupling as given in
Eq.~\eqref{decayop} together with Eq.~\eqref{lindblad} as function of tangle $\tau$ for different values of the parameter $q$.
For tangles larger than $1$, necessarily three Schmidt coefficients need to be positive,
and values larger than $4/3$ require at least four positive Schmidt coefficients.
The right-most inset of Fig.~\ref{lambda_decay} shows the a situation with significantly different decay times for the individual Schmidt-basis states.
A third and fourth basis state are occupied only for a value of $\tau$ above the threshold values $\tau=1$, and $\tau=4/3$.
This is different in the case of comparable life times of the individual basis states as depicted in the left-most inset.
Here, a third and fourth level are occupied already for significantly smaller values of $\tau$.
That is, even if a given value of tangle can be realized also with only two occupied levels, it is actually favorable
to occupy more levels, and thereby enhance the stability of entanglement.
This enhancement can be attributed to a larger occupation of the stable ground state,
as it can be seen in Fig.~\ref{lambda_decay}:
whenever an additional level gets occupied, there is a kink in the largest Schmidt coefficient
that shows how the decrease of the population of the stable levels is slowed down.

The situation is quite different in the case of decoherence that is displayed in Fig.~\ref{lambda_decoherence}.
For values of $q$ that are smaller than one, it is indeed favorable to occupy the levels $\ket{2}$ and $\ket{3}$, since the decoherence times
for superpositions of the states $\ket{2}\otimes\ket{2}$ and $\ket{3}\otimes\ket{3}$ are longest.
If $q$ is larger than one, the situation changes and superpositions of the states $\ket{0}\otimes\ket{0}$ and $\ket{1}\otimes\ket{1}$
are most stable so that there is a qualitative change between the two insets that correspond to the values $q=24/25$ and $q=26/25$.
A third and fourth level is occupied only above the threshold values $\tau=1$ and $\tau=4/3$ for all values of $q$.
However, the distribution of Schmidt coefficients at a given tangle $\tau>1$, rather strongly depends on $q$:
whereas the large Schmidt coefficients are of comparable size while there is single small Schmidt coefficient for $q\ll 1$ and for $q\gg 1$,
it is more favorable to have more broadly distributed Schmidt coefficients for $q\sim 1$.

\begin{figure}
\parbox{0.5\textwidth}{
\includegraphics[width=0.2\textwidth,angle=0]{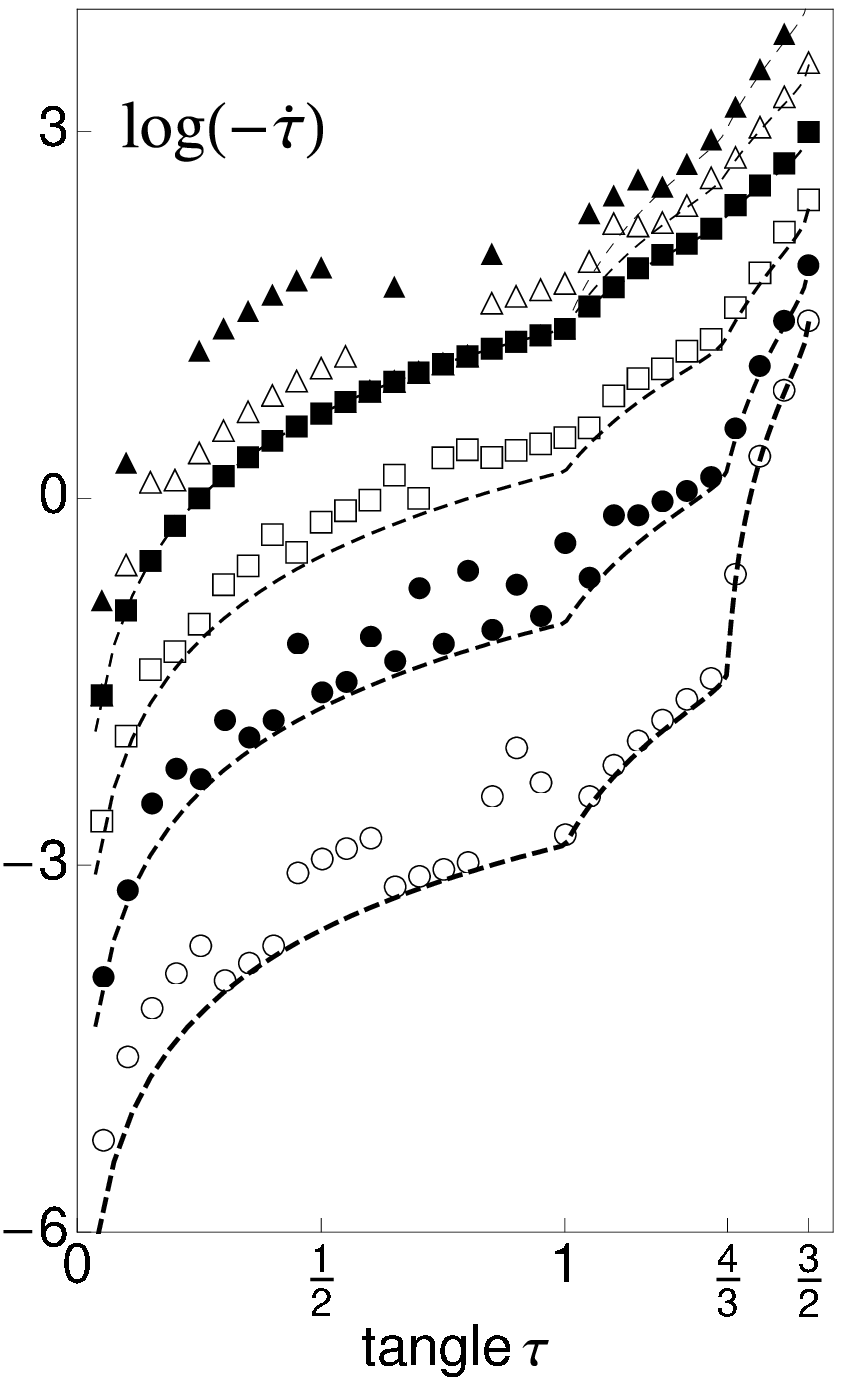}
\includegraphics[width=0.2\textwidth,angle=0]{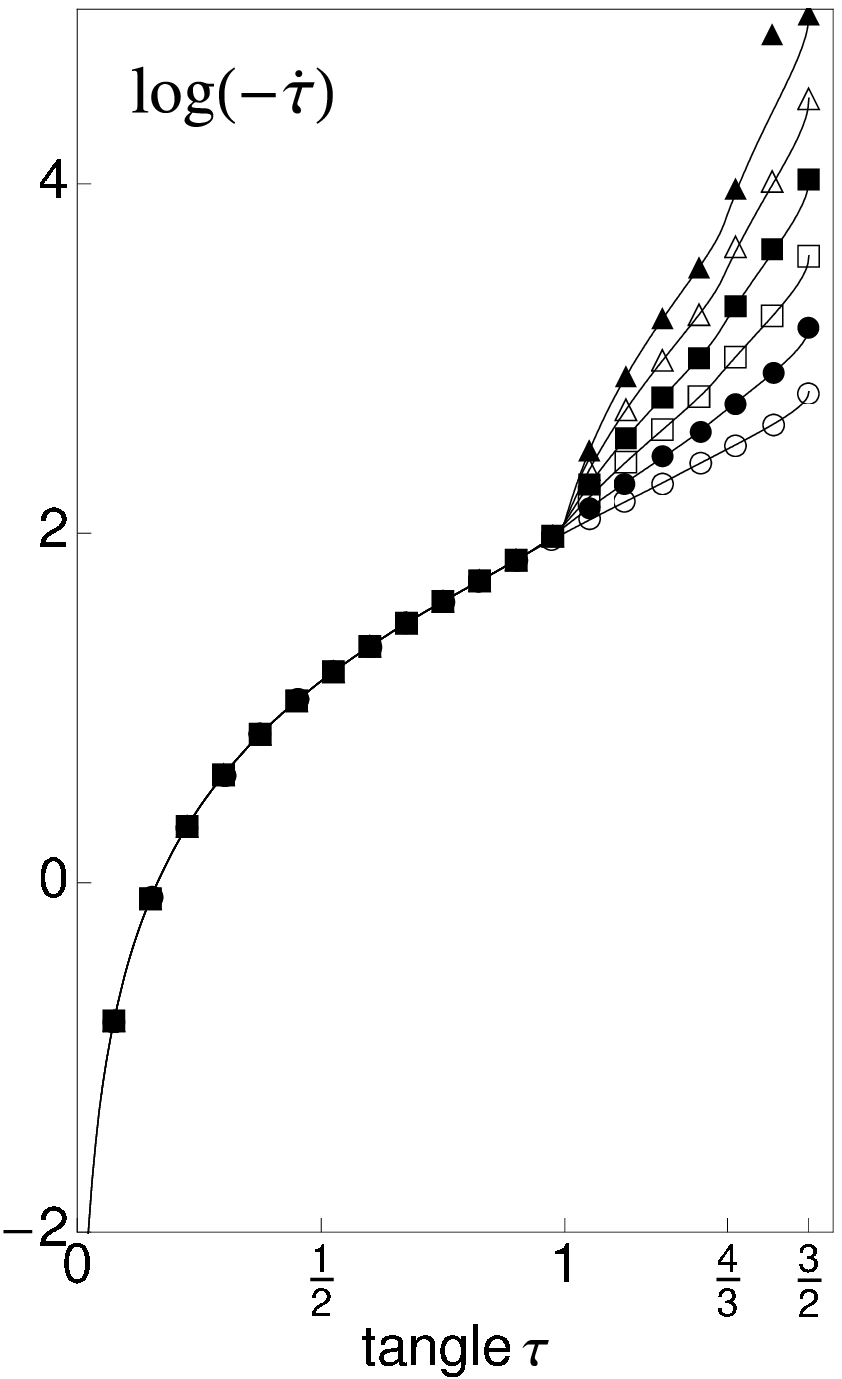}}
\caption{Comparison of numerically optimized temporal increment of tangle and optimized $\dot\tau$ with pointer bases as Schmidt bases:
the left inset corresponds to the situation of dephasing and the right inset to spontaneous decay.
The circles, squares and triangles display the numerical solutions for $q=1/4,1/2,3/4,1,5/4$ and $3/2$ (from bottom to top),
and the lines show the algebraically obtained solutions as function of the parameter $q$ defined in Eqs.~\eqref{decayop} and \eqref{decoherenceop}.}
\label{compare_alg_num}
\end{figure}

So far, we have been assuming that the optimal states have their Schmidt bases given in terms of the environment-induced pointer bases.
Fig.~\ref{compare_alg_num} shows a comparison of the algebraic solutions obtained under this assumption with numerical solutions,
where the optimization has been performed over general initial states.
The left inset displays the logarithm log$(-\dot\tau)$ of the temporal increment $\dot\tau$ of tangle for different values of $q$.
Apparently, all numerically obtained data points lie above the algebraic solutions that are shown as lines.
This implies that no improvement in the optimization can be obtained dropping the assumption on the Schmidt bases.
Besides that, there are also several data points that lie significantly above the algebraic solutions.
These points correspond to local maxima that pose a serious challenge to numerical optimizations,
that, however, are hardly a problem for our algebraic solutions.
The right inset of Fig.~\ref{compare_alg_num} shows the analogous situation for decay.
Here, the numerical optimization is significantly more reliable than in the case of decoherence,
and there is only one data point that corresponds to a local maximum.
Thus, the conjecture that Schmidt and pointer bases coincide is justified by the data shown.

\begin{figure}
\parbox{0.5\textwidth}{
\includegraphics[width=0.24\textwidth,angle=0]{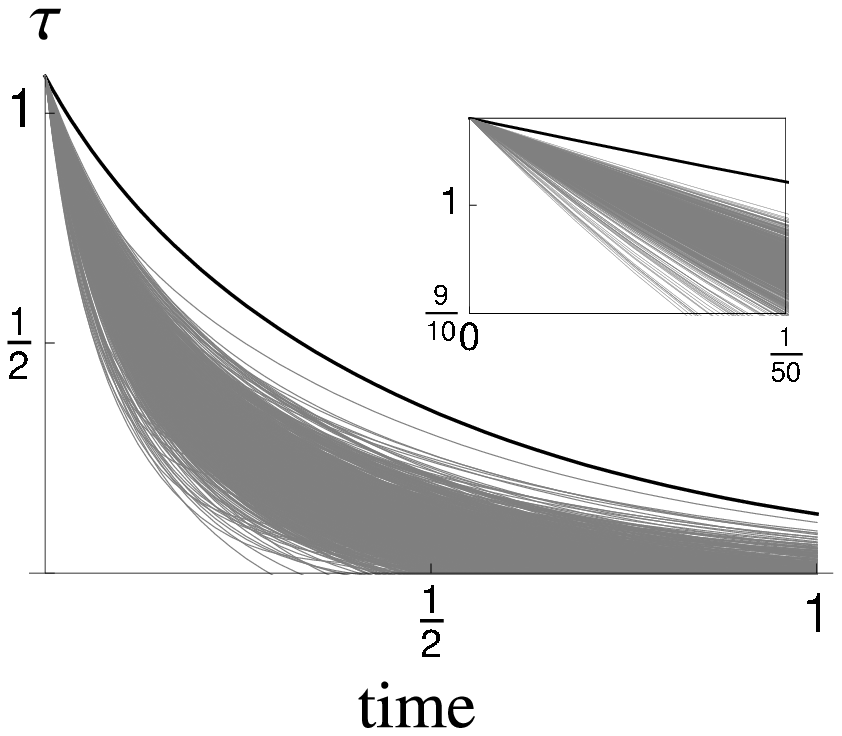}
\includegraphics[width=0.24\textwidth,angle=0]{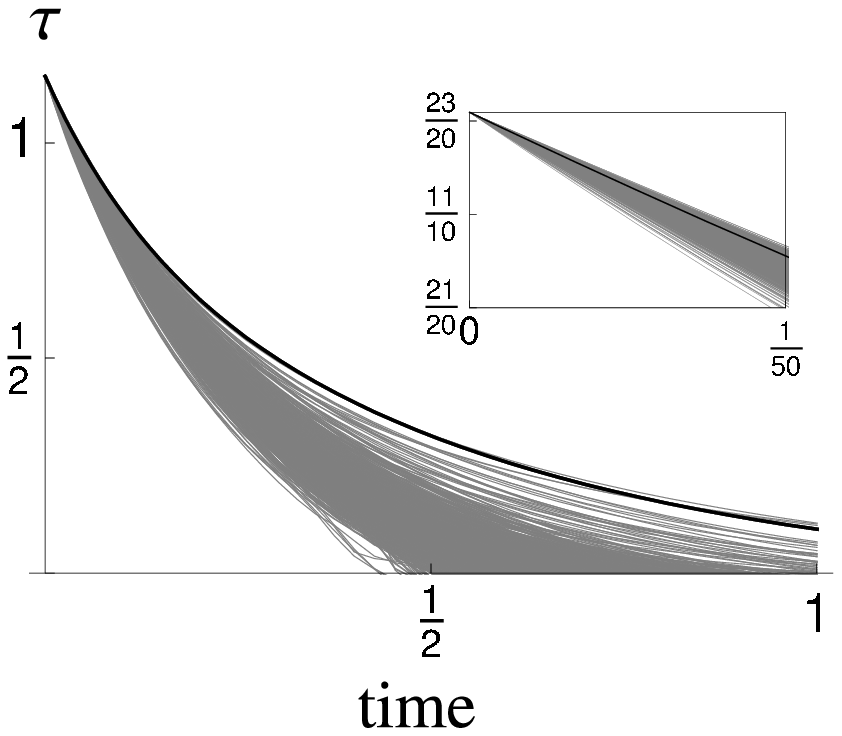}}
\caption{Time dependence of tangle for the obtained most robust states (black line) an $1000$ randomly chosen initial states with a given initial tangle $\tau_0$.
The left plot corresponds to decoherence (Eq.~\eqref{decoherenceop}) with $\tau_0=7/3$,
the right plot to decay (Eq.~\eqref{decayop}) with $\tau_0=8/3$.}
\label{compare_bounds}
\end{figure}

Finally, let us compare the optimal solutions that have been obtained with the approximation Eq.~\eqref{taudot}
with the time-dependence of state that is not subject to this approximation.
Here, we use the bounds \cite{mintert:167902,mintert:012336} that are known to provide a very accurate estimate of concurrence,
in particular for weakly mixed states, this is exactly the situation that we are dealing with here.
The black line in the left plot of Fig.~\ref{compare_bounds} shows the time evolution of $\tau$ for the state with initial tangle $\tau_0=7/3$
for decoherence with $q=1$.
The grey lines show the tangle for $1000$ randomly chosen states with the same initial tangle.
The tangle of the optimized case clearly decays significantly slower than that of the other states,
and the optimized states carry sizable entanglement at times where most other state have become separable.
The left plot of Fig.~\ref{compare_bounds} shows the analogous situation for decay with $q=1/10$ and initial states with initial tangle $\tau_0=8/3$.
Here, we have been choosing a very small value of $q$, since in this case the different excited states have comparable decay times
so that the formation of a pointer basis is not very pronounced.
Therefore, this example is a rather challenging test for our our approach that assumes the existence of such a basis.
In contrast to the left plot, there is no gap between the time evolutions of the random states and the optimized one,
and the inset that shows a zoom for the short time behavior indicates that there are a few states that have a slightly more robust entanglement
than the state that was found ideal.
However, the differences between the tangle of the optimized state and the tangle of the most robust random states is within the error margins
of the different estimates of mixed state tangle.

Another feature that strikes the eye here is the fact that in both insets of Fig.~\ref{compare_bounds} the tangle of the initially optimized states
turns out to be most robust over all times, although the optimization has only been performed for the initial time step.
Whereas this is a mere observation here, a more rigorous footing the generality of such behavior --
that would exceed the scope of the present letter by far -- will facilitate the search for robust mixed states tremendously.
In particular, prior observations that the dynamics of entanglement \cite{carvalho:190501} is often characterized very well
by the first infinitesimal time step strongly substantiate our conjecture that optimal mixed states are actually given by the decay products of initially pure states.

Here, we have been focussing on bipartite system.
The same ideas of maximizing the temporal increment of tangle can also be applied to multipartite systems for which Eq.~\eqref{tau}
can be generalized \cite{aolita:022308}.
The investigation for bipartite systems was facilitated with the Schmidt decomposition.
For general multipartite systems, however, there is not a simple generalization of this tool,
and there is an abundance of different classes of multipartite entangled states \cite{PhysRevA.62.062314,PhysRevA.65.052112}
that draws a very intransparent picture of entangled states.
Similarly to the bipartite case, where the Schmidt decomposition could be reproduced via the most robust states under very general situations of
environment coupling,
we expect that similar investigations on multipartite systems will allow to identify robust classes of entangled states
and provide a significantly more physical characterization of multi-partite states
than a mere distinction of mathematical classes can provide.

Stimulating discussions with Andreas Buchleitner and
financial support by the Alexander v. Humboldt foundation is gratefully acknowledged.

\bibliography{../../../../referenzen}

\begin{thebibliography}{22}
\expandafter\ifx\csname natexlab\endcsname\relax\def\natexlab#1{#1}\fi
\expandafter\ifx\csname bibnamefont\endcsname\relax
  \def\bibnamefont#1{#1}\fi
\expandafter\ifx\csname bibfnamefont\endcsname\relax
  \def\bibfnamefont#1{#1}\fi
\expandafter\ifx\csname citenamefont\endcsname\relax
  \def\citenamefont#1{#1}\fi
\expandafter\ifx\csname url\endcsname\relax
  \def\url#1{\texttt{#1}}\fi
\expandafter\ifx\csname urlprefix\endcsname\relax\def\urlprefix{URL }\fi
\providecommand{\bibinfo}[2]{#2}
\providecommand{\eprint}[2][]{\url{#2}}

\bibitem[{\citenamefont{Raussendorf and Briegel}(2001)}]{raussendorf:5188}
\bibinfo{author}{\bibfnamefont{R.}~\bibnamefont{Raussendorf}} \bibnamefont{and}
  \bibinfo{author}{\bibfnamefont{H.-J.} \bibnamefont{Briegel}},
  \bibinfo{journal}{Phys. Rev. Lett.} \textbf{\bibinfo{volume}{86}},
  \bibinfo{pages}{5188} (\bibinfo{year}{2001}).

\bibitem[{\citenamefont{D\"ur and Briegel}(2004)}]{dur:180403}
\bibinfo{author}{\bibfnamefont{W.}~\bibnamefont{D\"ur}} \bibnamefont{and}
  \bibinfo{author}{\bibfnamefont{H.-J.} \bibnamefont{Briegel}},
  \bibinfo{journal}{Phys. Rev. Lett.} \textbf{\bibinfo{volume}{92}},
  \bibinfo{eid}{180403} (\bibinfo{year}{2004}).

\bibitem[{\citenamefont{Carvalho et~al.}(2004)\citenamefont{Carvalho, Mintert,
  and Buchleitner}}]{carvalho:230501}
\bibinfo{author}{\bibfnamefont{A.~R.~R.} \bibnamefont{Carvalho}},
  \bibinfo{author}{\bibfnamefont{F.}~\bibnamefont{Mintert}}, \bibnamefont{and}
  \bibinfo{author}{\bibfnamefont{A.}~\bibnamefont{Buchleitner}},
  \bibinfo{journal}{Phys. Rev. Lett.} \textbf{\bibinfo{volume}{93}},
  \bibinfo{eid}{230501} (\bibinfo{year}{2004}).

\bibitem[{\citenamefont{Aolita et~al.}(2008{\natexlab{a}})\citenamefont{Aolita,
  Chaves, Cavalcanti, Ac\'{\i}n, and Davidovich}}]{aolita:080501}
\bibinfo{author}{\bibfnamefont{L.}~\bibnamefont{Aolita}},
  \bibinfo{author}{\bibfnamefont{R.}~\bibnamefont{Chaves}},
  \bibinfo{author}{\bibfnamefont{D.}~\bibnamefont{Cavalcanti}},
  \bibinfo{author}{\bibfnamefont{A.}~\bibnamefont{Ac\'{\i}n}},
  \bibnamefont{and}
  \bibinfo{author}{\bibfnamefont{L.}~\bibnamefont{Davidovich}},
  \bibinfo{journal}{Phys. Rev. Lett.} \textbf{\bibinfo{volume}{100}},
  \bibinfo{eid}{080501} (\bibinfo{year}{2008}{\natexlab{a}}).

\bibitem[{\citenamefont{Bennett et~al.}(1996)\citenamefont{Bennett, DiVincenzo,
  Smolin, and Wootters}}]{PhysRevA.54.3824}
\bibinfo{author}{\bibfnamefont{C.~H.} \bibnamefont{Bennett}},
  \bibinfo{author}{\bibfnamefont{D.~P.} \bibnamefont{DiVincenzo}},
  \bibinfo{author}{\bibfnamefont{J.~A.} \bibnamefont{Smolin}},
  \bibnamefont{and} \bibinfo{author}{\bibfnamefont{W.~K.}
  \bibnamefont{Wootters}}, \bibinfo{journal}{Phys. Rev. A}
  \textbf{\bibinfo{volume}{54}}, \bibinfo{pages}{3824} (\bibinfo{year}{1996}).

\bibitem[{\citenamefont{Rungta et~al.}(2001)\citenamefont{Rungta, Bu{\v z}ek,
  Caves, Hillery, and Milburn}}]{PhysRevA.64.042315}
\bibinfo{author}{\bibfnamefont{P.}~\bibnamefont{Rungta}},
  \bibinfo{author}{\bibfnamefont{V.}~\bibnamefont{Bu{\v z}ek}},
  \bibinfo{author}{\bibfnamefont{C.~M.} \bibnamefont{Caves}},
  \bibinfo{author}{\bibfnamefont{M.}~\bibnamefont{Hillery}}, \bibnamefont{and}
  \bibinfo{author}{\bibfnamefont{G.~J.} \bibnamefont{Milburn}},
  \bibinfo{journal}{Phys. Rev. A} \textbf{\bibinfo{volume}{64}},
  \bibinfo{pages}{042315} (\bibinfo{year}{2001}).

\bibitem[{\citenamefont{Mintert et~al.}(2005)\citenamefont{Mintert, Ku\'{s},
  and Buchleitner}}]{mintert:260502}
\bibinfo{author}{\bibfnamefont{F.}~\bibnamefont{Mintert}},
  \bibinfo{author}{\bibfnamefont{M.}~\bibnamefont{Ku\'{s}}}, \bibnamefont{and}
  \bibinfo{author}{\bibfnamefont{A.}~\bibnamefont{Buchleitner}},
  \bibinfo{journal}{Phys. Rev. Lett.} \textbf{\bibinfo{volume}{95}},
  \bibinfo{eid}{260502} (\bibinfo{year}{2005}).

\bibitem[{\citenamefont{Uhlmann}(1998)}]{uhl97}
\bibinfo{author}{\bibfnamefont{A.}~\bibnamefont{Uhlmann}},
  \bibinfo{journal}{Open Sys. \& Inf. Dyn.} \textbf{\bibinfo{volume}{5}},
  \bibinfo{pages}{209} (\bibinfo{year}{1998}).

\bibitem[{\citenamefont{Wootters}(1998)}]{PhysRevLett.80.2245}
\bibinfo{author}{\bibfnamefont{W.~K.} \bibnamefont{Wootters}},
  \bibinfo{journal}{Phys. Rev. Lett.} \textbf{\bibinfo{volume}{80}},
  \bibinfo{pages}{2245} (\bibinfo{year}{1998}).

\bibitem[{\citenamefont{Terhal and Vollbrecht}(2000)}]{ter00}
\bibinfo{author}{\bibfnamefont{B.~M.} \bibnamefont{Terhal}} \bibnamefont{and}
  \bibinfo{author}{\bibfnamefont{K.~G.~H.} \bibnamefont{Vollbrecht}},
  \bibinfo{journal}{Phys. Rev. Lett} \textbf{\bibinfo{volume}{85}},
  \bibinfo{pages}{2625} (\bibinfo{year}{2000}).

\bibitem[{\citenamefont{Mintert and Buchleitner}(2007)}]{mintert:140505}
\bibinfo{author}{\bibfnamefont{F.}~\bibnamefont{Mintert}} \bibnamefont{and}
  \bibinfo{author}{\bibfnamefont{A.}~\bibnamefont{Buchleitner}},
  \bibinfo{journal}{Phys. Rev. Lett.} \textbf{\bibinfo{volume}{98}},
  \bibinfo{eid}{140505} (\bibinfo{year}{2007}).

\bibitem[{\citenamefont{Lindblad}(1976)}]{Lindblad:1976fj}
\bibinfo{author}{\bibfnamefont{G.}~\bibnamefont{Lindblad}},
  \bibinfo{journal}{Comm. Math. Phys.} \textbf{\bibinfo{volume}{48}},
  \bibinfo{pages}{119} (\bibinfo{year}{1976}).

\bibitem[{\citenamefont{Hornberger}(2009)}]{klh}
\bibinfo{author}{\bibfnamefont{K.}~\bibnamefont{Hornberger}},
  \emph{\bibinfo{title}{Entanglement and Decoherence: Foundations and Modern
  Trends}} (\bibinfo{publisher}{Springer-Verlag, Berlin},
  \bibinfo{year}{2009}), chap. \bibinfo{chapter}{Introduction to decoherence
  theory}.

\bibitem[{\citenamefont{Nielsen and Chuang}(2000)}]{chuang00}
\bibinfo{author}{\bibfnamefont{M.}~\bibnamefont{Nielsen}} \bibnamefont{and}
  \bibinfo{author}{\bibfnamefont{I.}~\bibnamefont{Chuang}},
  \emph{\bibinfo{title}{Quantum Computation and Quantum Information}}
  (\bibinfo{publisher}{Cambridge University Press},
  \bibinfo{address}{Cambridge}, \bibinfo{year}{2000}).

\bibitem[{\citenamefont{Schmidt}(1907)}]{schmidt_deco}
\bibinfo{author}{\bibfnamefont{E.}~\bibnamefont{Schmidt}},
  \bibinfo{journal}{Math. Ann.} \textbf{\bibinfo{volume}{63}},
  \bibinfo{pages}{433} (\bibinfo{year}{1907}).

\bibitem[{\citenamefont{Adams and Loustaunau}(1994)}]{groebner}
\bibinfo{author}{\bibfnamefont{W.~W.} \bibnamefont{Adams}} \bibnamefont{and}
  \bibinfo{author}{\bibfnamefont{P.}~\bibnamefont{Loustaunau}},
  \emph{\bibinfo{title}{{A}n {I}ntroduction to {G}r\"obner {B}ases}}
  (\bibinfo{publisher}{American Mathematical Society}, \bibinfo{year}{1994}).

\bibitem[{\citenamefont{Mintert et~al.}(2004)\citenamefont{Mintert, Ku\'{s},
  and Buchleitner}}]{mintert:167902}
\bibinfo{author}{\bibfnamefont{F.}~\bibnamefont{Mintert}},
  \bibinfo{author}{\bibfnamefont{M.}~\bibnamefont{Ku\'{s}}}, \bibnamefont{and}
  \bibinfo{author}{\bibfnamefont{A.}~\bibnamefont{Buchleitner}},
  \bibinfo{journal}{Phys. Rev. Lett.} \textbf{\bibinfo{volume}{92}},
  \bibinfo{eid}{167902} (\bibinfo{year}{2004}).

\bibitem[{\citenamefont{Mintert and Buchleitner}(2005)}]{mintert:012336}
\bibinfo{author}{\bibfnamefont{F.}~\bibnamefont{Mintert}} \bibnamefont{and}
  \bibinfo{author}{\bibfnamefont{A.}~\bibnamefont{Buchleitner}},
  \bibinfo{journal}{Phys. Rev. A} \textbf{\bibinfo{volume}{72}},
  \bibinfo{eid}{012336} (\bibinfo{year}{2005}).

\bibitem[{\citenamefont{Carvalho et~al.}(2007)\citenamefont{Carvalho, Busse,
  Brodier, Viviescas, and Buchleitner}}]{carvalho:190501}
\bibinfo{author}{\bibfnamefont{A.~R.~R.} \bibnamefont{Carvalho}},
  \bibinfo{author}{\bibfnamefont{M.}~\bibnamefont{Busse}},
  \bibinfo{author}{\bibfnamefont{O.}~\bibnamefont{Brodier}},
  \bibinfo{author}{\bibfnamefont{C.}~\bibnamefont{Viviescas}},
  \bibnamefont{and}
  \bibinfo{author}{\bibfnamefont{A.}~\bibnamefont{Buchleitner}},
  \bibinfo{journal}{Phys. Rev. Lett.} \textbf{\bibinfo{volume}{98}},
  \bibinfo{eid}{190501} (\bibinfo{year}{2007}).

\bibitem[{\citenamefont{Aolita et~al.}(2008{\natexlab{b}})\citenamefont{Aolita,
  Buchleitner, and Mintert}}]{aolita:022308}
\bibinfo{author}{\bibfnamefont{L.}~\bibnamefont{Aolita}},
  \bibinfo{author}{\bibfnamefont{A.}~\bibnamefont{Buchleitner}},
  \bibnamefont{and} \bibinfo{author}{\bibfnamefont{F.}~\bibnamefont{Mintert}},
  \bibinfo{journal}{Phys. Rev. A} \textbf{\bibinfo{volume}{78}},
  \bibinfo{eid}{022308} (\bibinfo{year}{2008}{\natexlab{b}}).

\bibitem[{\citenamefont{D\"ur et~al.}(2000)\citenamefont{D\"ur, Vidal, and
  Cirac}}]{PhysRevA.62.062314}
\bibinfo{author}{\bibfnamefont{W.}~\bibnamefont{D\"ur}},
  \bibinfo{author}{\bibfnamefont{G.}~\bibnamefont{Vidal}}, \bibnamefont{and}
  \bibinfo{author}{\bibfnamefont{J.~I.} \bibnamefont{Cirac}},
  \bibinfo{journal}{Phys. Rev. A} \textbf{\bibinfo{volume}{62}},
  \bibinfo{pages}{062314} (\bibinfo{year}{2000}).

\bibitem[{\citenamefont{Verstraete et~al.}(2002)\citenamefont{Verstraete,
  Dehaene, De~Moor, and Verschelde}}]{PhysRevA.65.052112}
\bibinfo{author}{\bibfnamefont{F.}~\bibnamefont{Verstraete}},
  \bibinfo{author}{\bibfnamefont{J.}~\bibnamefont{Dehaene}},
  \bibinfo{author}{\bibfnamefont{B.}~\bibnamefont{De~Moor}}, \bibnamefont{and}
  \bibinfo{author}{\bibfnamefont{H.}~\bibnamefont{Verschelde}},
  \bibinfo{journal}{Phys. Rev. A} \textbf{\bibinfo{volume}{65}},
  \bibinfo{pages}{052112} (\bibinfo{year}{2002}).

\end{thebibliography}

\end{document}